\documentclass[10pt, twocolumn]{article}

\usepackage[utf8]{inputenc}

\usepackage{amsmath,amssymb,amsfonts}
\usepackage{xcolor}
\usepackage{amsthm}
\usepackage{xcolor}
\usepackage{physics}
\usepackage{comment}
\usepackage{graphicx}
\usepackage{subcaption}
\usepackage{algorithm, algpseudocode}
\usepackage{mathtools}
\usepackage{hyperref}
\usepackage[mathscr]{euscript}
\usepackage{xcolor}
\usepackage{svg}
\usepackage[style=ieee]{biblatex}
\usepackage{tcolorbox}
\usepackage[inline]{enumitem}

\DeclareMathOperator{\prob}{\mathsf{P}}
\DeclareMathOperator*{\argmax}{arg\,max}
\DeclareMathOperator*{\argmin}{arg\,min}

\newcommand{\EXPECT}{\mathrm{E}}
\newcommand{\COV}{\mathrm{Cov}}

\providecommand{\keywords}[1]{\textbf{Index terms---} #1}

\newsavebox{\smallmat} 

\begin{document}

\title{\textbf{The Economics of Privacy and Utility: Investment Strategies}}

\author{
Chandra~Sharma%
    \thanks{Chandra Sharma is a PhD student in the computer science department at Kansas State University,
KS, e-mail: ch1ndra@ksu.edu.},
George~Amariucai%
    \thanks{George Amariucai is an associate professor in the computer science department at Kansas State University, e-mail: amariucai@ksu.edu.},
and~Shuangqing~Wei%
    \thanks{Shuangqing Wei is a professor in the electrical and computer engineering department at Louisiana State University, e-mail: swei@lsu.edu.}
}

\date{}

\maketitle

\begin{abstract}
The inevitable leakage of privacy as a result of unrestrained disclosure of personal information has motivated extensive research on robust privacy-preserving mechanisms. However, existing research is mostly limited to solving the problem in a static setting with disregard for the privacy leakage over time. Unfortunately, this treatment of privacy is insufficient in practical settings where users continuously disclose their personal information over time resulting in an accumulated leakage of the users' sensitive information. In this paper, we consider privacy leakage over a finite time horizon and investigate optimal strategies to maximize the utility of the disclosed data while limiting the finite-horizon privacy leakage. We consider a simple privacy mechanism that involves compressing the user's data before each disclosure to meet the desired constraint on future privacy. We further motivate several algorithms to optimize the dynamic privacy-utility tradeoff and evaluate their performance via extensive synthetic performance tests.
\end{abstract}

\keywords{Dynamic privacy, Utility, Finite horizon, Kalman filter, Bellman equation.}

\section{Introduction}
The unprecedented growth of big-data applications suggests that there is a growing competition in the technological world to collect and harness tremendous amounts of user information. Tech companies and other online service providers are always seeking to enhance the quality of their products and services by collecting massive amounts of information from their user base. Users often disclose their personal information either by directly engaging with the service providers, such as in the case of social media and online shopping, or indirectly, and often inadvertently, by simply possessing IoT and other smart devices, such as location trackers. 

Users are often oblivious to the actual scale and nature of the information disclosure. The disclosed information often contains sensitive information about the users such as their current location, income, religious beliefs and sexual orientation. Further, service providers often share, and even sell, their customers' information with third parties, which makes protecting the users' private information ever so critical. In light of this, there have been increasing efforts to devise privacy-preserving mechanisms that make it difficult for external entities to infer a user's private information from the disclosed data. Such mechanisms protect the user's information often via means of distortion ~\cite{dwork2006our, sankar2013utility, he2013distortion, geng2015optimal, wang2016relation, kalantari2017information, liao2018privacy, kalantari2018robust} and/or compression ~\cite{diamantaras2016data, kung2017compressive, kung2017collaborative, kung2018compressive, 8768395} before disclosure. Unfortunately, the introduction of distortion (and/or compression) entails a loss of some useful information from the disclosed data which can otherwise be utilized by the service provider to provide a customized service to the user. The challenge, therefore, is to find an optimal tradeoff between protecting the user's privacy and enhancing the utility of the disclosed data.

Existing works that address the tradeoff between privacy and utility mostly treat privacy differently based on the context. In the context of data analysis, privacy is related to an adversary's ability to identify a user or the user's sensitive attributes in a database (for instance, \cite{alvim2011differential, yin2017location, yang2012differential, zhu2014correlated, wu2017game,he2014blowfish}), whereas in the context of information flow (over a noisy channel), privacy is related to an adversary's increase in knowledge about a user's sensitive attributes given some observable data (for instance, \cite{makhdoumi2013privacy, rajagopalan2011smart, sankar2013utility, sharma2021practical, braun2009quantitative, smith2009foundations, alvim2011differential,farokhi2017fisher, farokhi2017optimal, wang2018preserving,issa2016operational, liao2017hypothesis, li2018maximal, asoodeh2015maximal}). While existing works try to capture different notions of privacy and derive theoretical bounds on the privacy leakage, they do so in a static setting which either assumes that a user discloses their information only once, or it treats each disclosure of the user's information independently of the previous disclosures. This model of privacy does not accurately reflect real world settings in which users continuously disclose their personal information over time (as in social media) and each disclosure is temporally correlated with the previous disclosures. 
Therefore, privacy models under static settings are not only incomplete but also inaccurate for many practical applications. 

In this paper, we consider a dynamic model of privacy and capture privacy leakage over time, specifically focusing on the leakage at the end of a finite time horizon. Subject to the constraint on future privacy leakage, we investigate different strategies that yield different net utilities for the user. Notice that our privacy-utility tradeoff model is directly analogous to the investment problem from economics where a user seeks to maximize her rate of return over a finite time horizon by carefully choosing to invest a certain amount of money and spend the rest from her periodic income. Some examples of practical settings where our privacy model is useful are: \begin{enumerate*}
    \item A person aims to run for an office $n$ years from the present. The person seeks to cautiously regulate their social media usage in the meantime so that they can limit the amount of private information that can be inferred when they run for the office.
    \item A chemical plant is designed to undergo a complete overhaul after a certain number of years. In the meantime, the plant operator hires a third-party consultant, with expertise in data-based control strategies, with whom they decide to share some of the sensor data. At the same time, the operator aims to hide information about proprietary chemical processes at the time of the overhaul and therefore, looks to cautiously control the amount of shared information when the plant is in operation.
\end{enumerate*}

Overall, the main contributions of the paper are as follows:

\begin{enumerate}
    \item We formulate a novel privacy-utility tradeoff problem capturing the dynamics of privacy leakage over a finite time horizon. Our dynamic model of privacy also captures a practical setting in which a user's perception of their own privacy may change over time.
    \item Under our privacy-utility tradeoff model, we investigate different strategies that allow a user to maximize their net utility subject to certain privacy requirements.  
    \item We discuss challenges associated with finding optimal strategies for real world problems and motivate sub-optimal algorithms to solve the tradeoff problem.
    \item We extensively evaluate the performance of the sub-optimal algorithms on synthetic datasets and demonstrate that despite being sub-optimal, the proposed algorithms perform extremely well in achieving a good privacy-utility tradeoff.
    \item We formulate a simpler dynamic privacy problem that is computationally less intensive to solve but conserves the essence of the original problem.
\end{enumerate}

The rest of the paper is organized as follows. In Section \ref{section:related_work}, we briefly review existing works closely related to this paper. In Section \ref{section:problem_description}, we discuss the problem setting, explore the privacy and the utility requirements and motivate the finite-horizon privacy-utility tradeoff optimization problem. In Section \ref{section:formulation_mdp}, we formulate the finite-horizon privacy-utility tradeoff problem as a Markov Decision Process problem. In Section \ref{section:sub_optimal}, we discuss the challenges associated with solving the optimization problem and motivate sub-optimal algorithms. In Section \ref{section:estimated_leakage} we discuss a simpler problem model, highlight its advantages and present a computationally less intensive algorithm to solve the simplified problem. In Section \ref{section:simulations}, we evaluate the performance of all algorithms on synthetic datasets. Finally, in Section \ref{section:conclusion}, we sum up our work with closing remarks.

\section{Related Work} \label{section:related_work}
The problem of optimizing the privacy-utility tradeoff in a static setting has been widely studied under different notions of privacy. Existing works mostly focus on precisely defining privacy based on the context and deriving bounds on the privacy leakage using metrics such as KL-divergence \cite{li2009tradeoff}, {Differential Privacy} \cite{alvim2011differential, yin2017location, yang2012differential}, {Correlated Differential Privacy} ~\cite{zhu2014correlated, wu2017game}, {Blowfish Privacy} \cite{he2014blowfish}, {Mutual Information} \cite{makhdoumi2013privacy, rajagopalan2011smart, sankar2013utility, kalantari2017information, basciftci2016privacy}, {Changes in min-entropy} \cite{braun2009quantitative, smith2009foundations, alvim2011differential}, {Fisher Information} \cite{farokhi2017fisher, farokhi2017optimal, wang2018preserving}, {Maximal Leakage} \cite{issa2016operational, liao2017hypothesis}, $\chi^2$-information \cite{wang2017estimation} and {Maximal Correlation} \cite{li2018maximal, asoodeh2015maximal, makhdoumi2013privacy}). However, most existing works do not model a continuous and correlated disclosure of information and therefore, do not capture the privacy leakage over time.

There are but a few works that model the continuous disclosure of a user's information and capture the temporal correlation between subsequent disclosures. In \cite{erdogdu2015privacy}, the authors investigate the privacy leakage resulting from a continuous release of a time-series data that is correlated, both spatially and temporally, with a user's sensitive data (also considered to be time-series but non-disclosable). The privacy mechanism seeks to distort the time series data before each disclosure to impede inference attacks on the sensitive data while preserving the utility of the disclosed data. This model seeks to limit the privacy leakage at the present time using the information from the past disclosures and by carefully regulating the current disclosure (other similar models can be found in \cite{zhou2009continuous, dwork2010differential}). In contrast, our model seeks to limit the privacy leakage in the future using the information from the past disclosures and by carefully regulating the present and all future disclosures. Due to the uncertainties around the future observations, our model is more complex, but also more general, and easily simplifies to a model similar to that in \cite{erdogdu2015privacy} under a particular instantiation (namely, $n = k$ where $n$ represents the finite time horizon and $k$ represents the current time step).

Recently, researchers have started to explore the privacy issues in a dynamic setting with regard to future privacy leakage. In \cite{8462600}, the authors investigate the privacy issues related with the continuous disclosure of sensor measurements containing some private and some public information. They formulate the problem as a filtering problem in which they seek to find the optimal compression that maximizes the variance of the estimation error associated with the estimation of the private data while minimizing the variance of the estimation error associated with the estimation of the public data. Under their model, they investigate the privacy-utility tradeoff at the current time step and one time step into the future. 
The same work is further extended in \cite{8768395} where the authors investigate the tradeoff multiple time steps into the future. In their formulation, they make predictions about the system's future state using the observations available up until the current time step. However, since future observations are discounted in making the prediction, the predicted state can be quite off from the actual future state. Further, the constraint is formulated in terms of the predicted future privacy leakage instead of the actual leakage. Therefore, while this model can be useful, it is not quite complete as it does not accurately reflect the actual future privacy leakage. In many practical applications, due to the inherent uncertainty about the future observations, no strategy can satisfy a non-trivial privacy constraint with certainty. A comprehensive analysis of the privacy-utility tradeoff, therefore, requires the investigation of the probability of privacy outage, i.e. the probability that the privacy constraint is not satisfied in the future.

\vspace{-2mm}

\section{Problem Description} \label{section:problem_description}
\subsection{Problem Setup}
We consider a setting where a privacy-aware user seeks to cautiously disclose her personal information to a data analyst over a finite time horizon. The objective of the user is to maximize her instantaneous utilities, which the data analyst provides by extracting useful information from the disclosed information at each time step, while limiting the amount of leakage about her sensitive information at the end of the finite time horizon. In contrast to the static setting which models the information disclosure at a single time step, the dynamic setting under consideration models the incremental disclosure of information at every time step until the end of the finite time horizon. The solution to this dynamic privacy problem involves finding an optimal strategy that maximizes the sum of instantaneous utilities while ensuring that the privacy leakage at the end of the finite horizon remains below a pre-specified threshold with high probability. 


In the dynamic privacy setting, we assume that each user has a set of features, represented by the random vector $X$, which evolves over time. We use the subscript $k$ to denote the feature vector at time step $k$. At any given time step $k$, the feature vector consists of the user's private features represented by the random vector $X^p_k$ and the user's public features represented by the random vector $X^u_k$ such that $X_k = X^p_k \cup X^u_k$. We consider a general setting where $X^p_k$ and $X^u_k$ are correlated. In many real world settings, the user's observations of her own feature vectors are only available as noisy measurements (for instance, heart-rate readings from a smart watch). To model this, we assume that the true values of $X_k$ (and consequently, $X^p_k$ and $X^u_k$) may not be directly observable; instead, there is an observable process $Z_k$ that carries information about $X_k$ such that $Z_k = f(X_k)$. The user's instantaneous privacy and utility is directly associated with $X^p_k$ and $X^u_k$, respectively. Next, we assume that the user is willing to disclose $Z_k$ in return for some utility. However, as $Z_k$ contains information about both $X^p_k$ and $X^u_k$, disclosing $Z_k$ inevitably leaks some information about $X^p_k$, and this leakage carries over to the future time-steps which the user seeks to avoid. To address this, we consider a privacy mechanism which perturbs $Z_k$ before disclosure. The privacy mechanism involves transforming the entire observation vector, $Z_k$, into a lower-dimensional noisy vector, $\tilde{Z_k}$. The transformation is essentially non-invertible and therefore, certain information about $Z_k$ (and consequently, $X_k$) is lost due to the transformation. An ideal transformation function maximizes the information loss regarding $X^p_k$ while minimizing the information loss regarding $X^u_k$. However, due to the correlation between $X^p$ and $X^u$, this may not always be possible.

A Linear Dynamical System (LDS), which is a continuous state-space generalization of a Hidden Markov Model, can be used to model the evolution of the user's feature vectors over time as well as the observation process. 
LDS has been widely used to model the underlying system in the context of privacy-preserving information disclosure \cite{8462600, 8768395, koufogiannis2017differential, han2018privacy, sugiura2021bayesian}. 
We consider a first-order LDS model in which $X_k$ evolves according to the linear equation:
\begin{align}
    X_{k} = F_k X_{k-1} + W_k,  \label{eqn:lds_state_evolution}
\end{align}
where $F_k$ is the \textit{state-transition matrix} and $W_k$ is the zero-mean Gaussian process noise with covariance $Q_k$. Similarly, the observation process can be represented by the linear equation:
\begin{align}
    Z_{k} = H_k X_{k} + V_k,    \label{eqn:lds_measurement}
\end{align}
where $H_k$ denotes the \textit{observation matrix} and $V_k$ denotes the zero-mean Gaussian measurement noise with covariance $R_k$. We consider both $F_k$ and $H_k$ to be full-ranked square matrices and assume that all system parameters are publicly known. For quick reference, the description of all system parameters, system vectors and other symbols used throughout this paper can be found in Table \ref{table:symbol}.  

\begin{table}[!ht]
    \caption{Description of Some Commonly Used Symbols and Notations}
    \centering
    \begin{tabular}{| p{0.12\linewidth} | p{0.8\linewidth} |}
    \hline
        \textbf{Symbol} & \textbf{Description} \\
        \hline
        $n$ & Finite time horizon \\
        \hline
        $N_p$ & Number of private features \\
        \hline
        $N_u$ & Number of utility features \\
        \hline
        $N$ & Total number of features ($N = N_p + N_u$) \\
        \hline
        $M$ & Compression size ($M < N$) \\
        \hline
        $X_k$ & Random vector representing the user's features ($X_k \in \mathbb{R}^{N \times 1}$)\\
        \hline
        $Z_k$ & Observation vector ($Z_k \in \mathbb{R}^{N \times 1}$)\\
        \hline
        $\tilde{Z_k}$ & Compressed observation vector ($\tilde{Z}_k \in \mathbb{R}^{M \times 1}$)\\
        \hline
        $F_k$ & State-transition matrix ($F_k \in \mathbb{R}^{N \times N}$)\\
        \hline
        $H_k$ & Observation matrix ($H_k \in \mathbb{R}^{N \times N}$)\\
        \hline
        $Q_k$ & The covariance of the process noise ($Q_k \in \mathbb{R}^{N \times N}$) \\
        \hline
        $R_k$ & The covariance of the measurement noise ($R_k \in \mathbb{R}^{N \times N}$) \\
        \hline
        $C_k$ & Compression matrix ($C_k \in \mathbb{R}^{N \times M}$)\\
        \hline
        $X^p_k$ & Random vector representing the user's true private feature at time $k$ ($X^p_k \in \mathbb{R}^{N_p \times 1}$) \\
        \hline
        $X^u_k$ & Random vector representing the user's true public (utility) feature at time $k$ ($X^u_k \in \mathbb{R}^{N_u \times 1}$) \\
        \hline
        $\widehat{X}^{Z,p}_{k|j}$ & Data owner's estimation of $X^p_k$ using observations $Z_1, Z_2, \cdots, Z_j$ ($k \geq j$) \\
        \hline
        $\widehat{X}^{Z,u}_{k|j}$ & Data owner's estimation of $X^u_k$ using observations $Z_1, Z_2, \cdots, Z_j$ ($k \geq j$) \\
        \hline
        $\widehat{X}^{\tilde{Z},p}_{k|j}$ & Adversary's estimation of $X^p_k$ using observations $\tilde{Z}_1, \tilde{Z}_2, \cdots, \tilde{Z}_j$ ($k \geq j$) \\
        \hline
        $\widehat{X}^{\tilde{Z},u}_{k|j}$ & Data analyst's estimation of $X^u_k$ using observations $\tilde{Z}_1, \tilde{Z}_2, \cdots, \tilde{Z}_j$ ($k \geq j$) \\
        \hline
        $P^Z_{k|k}$ & Error covariance associated with the data owner's estimate of $X_k$  \\
        \hline
        $P^{\tilde{Z}}_{k|k}$ & Error covariance associated with the data analyst and adversary's estimate of $X_k$  \\
        \hline
        $\prob (X|Y)$ & The conditional probability of X given Y \\
    \hline
    \end{tabular}
    \label{table:symbol}
\end{table}

\begin{figure}
    \includegraphics[width = \linewidth]{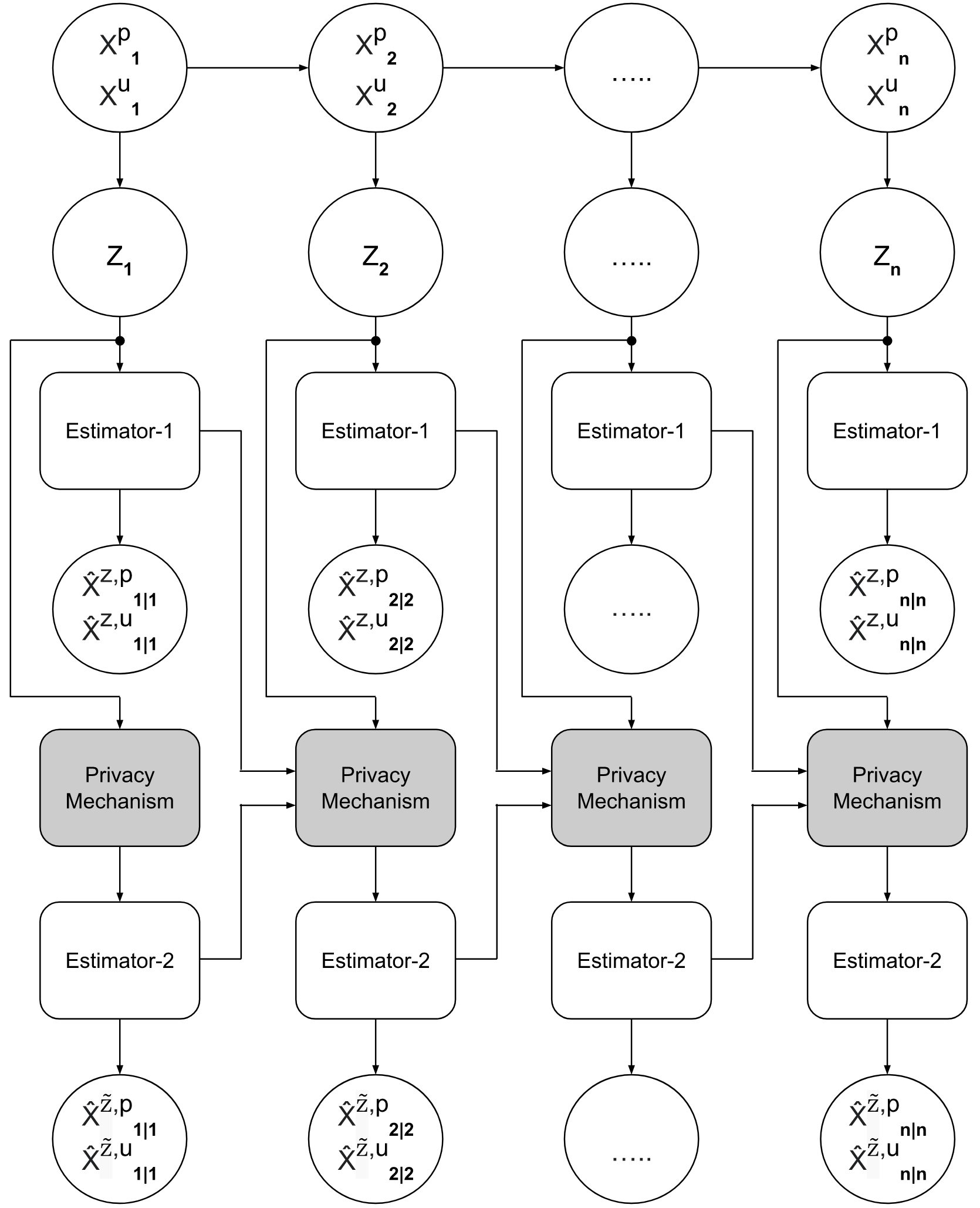}
    \caption{The dynamics of the finite-horizon privacy-utility tradeoff problem.}
    \label{figure:estimation}
\end{figure}

The privacy mechanism involves mapping the observation vector $Z_k$ to a lower-dimensional vector $\tilde{Z}_k$ using a compression matrix $C_k$ such that $\tilde{Z}_k = C_k^T Z_k$. We assume that an adversary has complete knowledge about the system dynamics as well as the privacy mechanism. The goal of the data-owner (user) is to prudently select the compression matrices, $C_1, C_2, \cdots C_n$, that maximize the sum of instantaneous utilities while limiting the amount of information leaked about the private features at the end of the finite time horizon, $n$. Note that the sequence $C_1, C_2, \cdots, C_n$ constitutes the \textit{strategy} for the data-owner. The data analyst is tasked with inferring $X^u_k$ from the disclosed sequence $\tilde{Z}_1, \tilde{Z}_2, \cdots, \tilde{Z}_k$ at each time step $k$ while the future adversary seeks to infer $X^p_n$ from the disclosed sequence $\tilde{Z}_1, \tilde{Z}_2, \cdots, \tilde{Z}_n$. The problem, therefore, naturally relates to an estimation problem.
The dynamics of the problem are depicted in Figure \ref{figure:estimation}.

Before discussing the formal models of privacy and utility, we first focus on the problem of estimating the latent system states, $X^p_k$ and $X^u_k$, given a series of observations. This estimation problem can be solved using the \textit{Kalman filter} which is an optimal linear filter in terms of minimizing the \textit{Mean Squared Error} of the estimates \cite{welch2020kalman}. Estimation using the Kalman filter involves two steps: the prediction step in which the system states are predicted a priori and the update step in which the current measurements/observations are incorporated to update the state estimates. Formally, the Kalman filter for the LDS represented by (\ref{eqn:lds_state_evolution}) and (\ref{eqn:lds_measurement}) can be expressed as \cite{welch2020kalman}:

\vspace{2mm}
\textit{Prediction step:}
\begin{align*}
    \widehat{x}_{k|k-1} &= F_k \widehat{x}_{k-1|k-1} \\
    P_{k|k-1} &= F_k P_{k-1|k-1} F_k^T
\end{align*}

\textit{Update step:}
\begin{align*}
    \widehat{x}_{k|k} &= \widehat{x}_{k|k-1} + K_k(Z_k - H_k \widehat{x}_{k|k-1} ) \\
    P_{k|k} &= P_{k|k-1} - K_k H_k P_{k|k-1},
\end{align*}
where $\widehat{x}_{k|k-1}$ is the a priori estimate of $X_k$ given the observations up to time $k-1$, $\widehat{x}_{k|k}$ is the a posteriori estimate of $X_k$ given the observations up to time $k$, $P_{k|k-1}$ is the a priori error covariance of the estimate $\widehat{x}_{k|k-1}$, and $P_{k|k}$ is the a posteriori error covariance of the estimate $\widehat{x}_{k|k}$. The Kalman gain, $K_k$, is given by $K_k = P_{k|k-1} H_k^T (H_k P_{k|k-1} H_k^T + R_k)^{-1}$.

\subsection{Privacy and Utility Requirements}
Let $\widehat{X}^{Z,u}_{k|k}$ and $\widehat{X}^{Z,p}_{k|k}$ represent the data owner's estimate of $X^u_k$ and $X^p_k$, respectively, given the series of observations, $Z_1, Z_2, \cdots, Z_k$. Similarly, let $\widehat{X}^{\tilde{Z},u}_{k|k}$ and $\widehat{X}^{\tilde{Z},p}_{k|k}$ represent the data analyst's estimate of $X^u_k$ and the adversary's estimate of $X^p_k$, respectively, given the series of observations, $\tilde{Z}_1, \tilde{Z}_2, \cdots, \tilde{Z}_k$. Also, let $d(X,Y)$ denote some distance function that measures the distance between random vectors X and Y. An example of the distance function is the $L^2$-norm. From the utility point of view, it is desirable that $d(\widehat{X}^{\tilde{Z},u}_{k|k}, \widehat{X}^{Z,u}_{k|k})$ is as small as possible for all $k$. A zero distance between the estimates, $\widehat{X}^{\tilde{Z},u}_{k|k}$ and $\widehat{X}^{Z,u}_{k|k}$, is achievable if the data owner discloses her true observations, $Z_1, Z_2, \cdots, Z_k$, with no privacy mechanisms. Similarly, from the future privacy point of view, it is desirable that $d(\widehat{X}^{Z,p}_{n|n}, \widehat{X}^{\tilde{Z},p}_{n|n} )$ is as large as possible. However, due to the correlation between $X^p_k$ and $X^u_k$, in general, it is not feasible to both minimize the instantaneous utility losses and maximize the perceived future privacy. The problem, therefore, naturally manifests as a privacy-utility trade-off optimization problem. Intuitively, the optimization problem involves finding an optimal strategy that minimizes the sum of instantaneous utility losses, $\sum_{k = 1}^{n} d(\widehat{X}^{\tilde{Z},u}_{k|k}, \widehat{X}^{Z,u}_{k|k})$, under the constraint that the privacy leakage at the end of the finite horizon, $\frac{1}{d(\widehat{X}^{\tilde{Z},p}_{n|n}, \widehat{X}^{Z,p}_{n|n})}$, must not exceed a pre-specified threshold $\frac{1}{\delta}$. 
Formulating the optimization problem, however, exposes several challenges. For one, $\sum_{k = 1}^{n} d(\widehat{X}^{\tilde{Z},u}_{k|k}, \widehat{X}^{Z,u}_{k|k})$ and $d(\widehat{X}^{\tilde{Z},p}_{n|n}, \widehat{X}^{Z,p}_{n|n})$ are both random variables due to the uncertainties in the future observations, $Z_{k+1}, Z_{k+2}, \cdots, Z_n$. Further, without the knowledge of the future observations, it is difficult to devise an optimal strategy that satisfies the constraint on future privacy leakage. In fact, given the Gaussian assumption for both the process noise and the measurement noise, it may not even be possible to ensure a non-trivial constraint on the future privacy leakage using any feasible sequence of actions, $C_1, C_2, \cdots, C_n$; it is, therefore, more appropriate to characterize strategies in terms of the probability of privacy outage, $\prob(d(\widehat{X}^{\tilde{Z},p}_{n|n}, \widehat{X}^{Z,p}_{n|n}) < \delta)$, in addition to the total utility loss. The probability of privacy outage reflects the probability that the privacy constraint is not satisfied in the future.

\section{Formulation as a Markov Decision Process Problem} \label{section:formulation_mdp}

The finite horizon privacy-utility trade-off problem fits nicely with a Markov Decision Process (MDP). In a discrete time continuous state MDP model, at every time step $k$, an agent observes the current state of some Markov process $S_k$, takes an action $a_k$ and receives a reward $R_k$. The state of the Markov process at time step $k$, in general, depends on the state and the action at time step $k-1$ and some stochastic process noise $\omega_{k}$. The reward received by the agent at time step $k$ depends on the current state of the Markov process, the current action taken by the agent and the next state of the Markov process.

Formally, a discrete time continuous state continuous action Markov Decision Process is a tuple $(\mathscr{S}, \mathscr{A}, P, R, \gamma)$ where $\mathscr{S} \in \mathbb{R}^J$ represents the state space, $\mathscr{A} \in \mathbb{R}^{L}$ represents the action space, $P: \mathscr{S} \times \mathscr{A} \times \mathscr{S} \rightarrow [0,1] $ represents the state-transition function such that $P\left(s_k | a_k, s_{k+1}\right)$ gives the probability of transitioning to the next state $s_{k+1}$ from the current state $s_k$ by taking an action $a_k$. Let the random vectors $S_k$, $A_k$ and $S_{k+1}$ represent the current state, the current action and the next state, respectively. As the state space is continuous, $P$ is specified as a probability density function such that $\int_{\Psi} P\left(s_k | a_k, s_{k+1}\right) \mathrm{d}s_{k+1}= \prob(S_{k+1} \in \Psi | S_k = s_k, A_k = a_k)$, where $\Psi \subseteq \mathscr{S}$, with $\mathscr{S}$ the space of $S_k$.
Similarly, $R : \mathscr{S} \times \mathscr{A} \times \mathscr{S} \rightarrow \mathbb{R}$ represents the reward function such that $R_k(s_k, a_k, s_{k+1})$ gives the reward received by the agent at time step $k$ by taking an action $a_k$ when the current and the next states of the Markov process are $s_k$ and $s_{k+1}$, respectively. The discount factor $\gamma \in [0,1]$ captures how the agent values her future rewards compared to her current reward -- if $\gamma = 1$, the agent values all her future rewards equally to her current reward and if  $\gamma = 0$, the agent only values her current reward and disregards all her future rewards.

The goal of the agent is to maximize the expected sum of her current and future discounted rewards, $\EXPECT[\sum_k \gamma^k R_k(S_k, A_k, S_{k+1})]$. The agent seeks to find the optimal sequence of actions that allows her to optimize the expected sum of discounted rewards. In this regard, it is useful to define a function, called the optimal state-value function, that provides a measure of the maximum achievable sum of expected rewards from a particular state. Let $V_k^\ast(s_k)$ denote the optimal state-value function at time step $k$ given the current state $s_k$. Then, the optimal state-value function can be  written as a recursive equation using Bellman's Principle of Optimality as:
\begin{align*}
    V_k^\ast(s_k) = \max_{a_k} \int_{\mathscr{S}} P\left(s_k | a_k, s_{k+1}\right) \big(R_k(s_k, a_k, s_{k+1}) + \\
    \gamma V_{k+1}^\ast(s_{k+1}) \big) \mathrm{d}s_{k+1}.
\end{align*}
The Bellman equation formulation offers a dynamic programming approach to solve the resulting optimization problem.

The finite-horizon privacy-utility problem can be directly translated to a finite-horizon discrete time continuous state continuous action MDP problem. Recall that in the finite-horizon privacy setting, the user seeks to find an optimal sequence of actions that allows her to maximize the sum of the instantaneous utilities while ensuring that the privacy leakage at the end of the finite horizon remains below a pre-specified threshold with high probability. This is inherently a decision problem that incorporates a meaningful notion of reward captured as the expected sum of instantaneous utilities and future privacy leakage.

Let $\widehat{X}^Z_{k|k}$ represent the data owner's estimate of $X$, given the series of observations, $Z_1, Z_2, \cdots, Z_k$ and $P^Z_{k|k}$ represent the error covariance associated with the estimate\footnote{$P^Z_{k|k}$ is not a function of $Z_k$. The superscript $Z$ is used as a convention to imply that the symbol being defined directly concerns the data owner, who has observations $\{Z_k\}$, rather than the adversary or data analyst, who have observations $\{\tilde{Z}_k\}$.}. Similarly, let $\widehat{X}^{\tilde{Z}}_{k|k}$ represent the data analyst's (or adversary's) estimate of $X$, given the series of observations, $\tilde{Z}_1, \tilde{Z}_2, \cdots, \tilde{Z}_k$ and $P^{\tilde{Z}}_{k|k}$ represent the error covariance associated with the estimate.
Now, define
\begin{align*}
   &d^u(k,k) \triangleq d(\widehat{X}^{Z,u}_{k|k}, \widehat{X}^{\tilde{Z},u}_{k|k}),
   \\
   &d^p(j,k) \triangleq d(\widehat{X}^{Z,p}_{j|k}, \widehat{X}^{\tilde{Z},p}_{j|k}) \quad (n \geq j \geq k),
   \\
   &S_k \triangleq \{Z_k, \widehat{X}^Z_{k-1|k-1}, \widehat{X}^{\tilde{Z}}_{k-1|k-1}, P^Z_{k-1|k-1}, P^{\tilde{Z}}_{k-1|k-1}\},
\end{align*}
where $S_k$ represents the state of the MDP (which is fundamentally different from the state of the LDS, $X_k$). The state variables in $S_k$ can be recursively computed using the following sequence of Kalman filter equations:
\begin{align*}
    &\widehat{X}^Z_{k|k-1} = F_k \widehat{X}^Z_{k-1|k-1},
    \\
    &\widehat{X}^{\tilde{Z}}_{k|k-1} = F_k \widehat{X}^{\tilde{Z}}_{k-1|k-1}, 
    \\
    &P^Z_{k|k-1} = F_k P^Z_{k-1|k-1} F_k^T + Q_k,
    \\
    &P^{\tilde{Z}}_{k|k-1} = F_k P^{\tilde{Z}}_{k-1|k-1} F_k^T + Q_k,
    \\
    &K^Z_k = P^Z_{k|k-1} H_k^T (H_k P^Z_{k|k-1} H_k^T + R_k)^{-1},
    \\
    &K^{\tilde{Z}}_k = P^{\tilde{Z}}_{k|k-1} H_k^T C_k (C_k^T H_k P^{\tilde{Z}}_{k|k-1} H_k^T C_k + C_k^T R_k C_k)^{-1},
    \\
    &\widehat{X}^Z_{k|k} = \widehat{X}^Z_{k|k-1} + K^Z_k (Z_k - H_k \widehat{X}^Z_{k|k-1}), 
    \\
    &\widehat{X}^{\tilde{Z}}_{k|k} = \widehat{X}^{\tilde{Z}}_{k|k-1} + K^{\tilde{Z}}_k (C_k^T Z_k - C_k^T H_k \widehat{X}^{\tilde{Z}}_{k|k-1}) ,
    \\
    &P^Z_{k|k} = P^Z_{k|k-1} - K^Z_k H_k P^Z_{k|k-1},
    \\
    &P^{\tilde{Z}}_{k|k} = P^{\tilde{Z}}_{k|k-1} - K^{\tilde{Z}}_k C_k^T H_k P^{\tilde{Z}}_{k|k-1}.
\end{align*}

Initially, $\widehat{X}^Z_{0|0}$ = $\widehat{X}^{\tilde{Z}}_{0|0} = \EXPECT[X_0]$ and $P^Z_{0|0}= P^{\tilde{Z}}_{0|0} = \mathrm{Cov}(X_0)$.

\vspace{2mm}
We now define the reward function, $R_k$, as
\begin{align}
    &R_k(S_k, C_k, S_{k+1}) = \nonumber\\ 
        &\begin{cases}    
            \alpha\big( d^p(n,k+1) - d^p(n,k) \big) - d^u(k,k) & \text{when $k < n$,} \\
            \alpha d^p(n,k) - d^u(k,k) & \text{when $k = n$},
        \end{cases}
\end{align}
where $\alpha$ is the privacy-utility tradeoff parameter.

At any given time $k$, the user's goal is to chose an action $C_k = f(S_k)$ that allows her to maximize the expected sum of the current and future rewards, $\EXPECT[\sum_{t=k}^{n} R_t(S_t, C_t, S_{t+1})]$.

Let ${\bf{C}}^\ast = \{ C_1^\ast, C_2^\ast , \cdots, C_n^\ast \}$ be the set of optimal actions. Notice that at the beginning of the finite time horizon, the sum of the rewards can be expressed as
\begin{align*}
    &\sum_{k=0}^{n} R_k(S_k, C_k, S_{k+1}) \\
        &= \alpha\big( 2d^p(n,n) - d^p(n,0) \big) - \sum_{k = 0}^{n}d^u(k,k).
\end{align*}


Since $d^P(n,0)$ and $d^u(0,0)$ are both zero (which follows from the assumption that $\widehat{X}^Z_{0|0} = \widehat{X}^{\tilde{Z}}_{0|0}$ and $P^Z_{0|0} = P^{\tilde{Z}}_{0|0}$), substituting $2\alpha = \beta$, we get
\begin{align}
    \sum_{k=0}^{n} R_k(S_k, C_k, S_{k+1}) 
        = \beta d^p(n,n) - \sum_{k = 1}^{n}d^u(k,k). 
        \label{expr:reward}
\end{align}

As the reader may have noticed by now, the reward function is defined such that the sum of rewards captures both the privacy and the utility aspects of the problem in a single expression given in (\ref{expr:reward}). The value of $\gamma$ is taken to be $1$ for the same reason. Note that the parameter $\beta$ in (\ref{expr:reward}) directly relates to the probability of privacy outage $\prob(d(\widehat{X}^{\tilde{Z},p}_{n|n}, \widehat{X}^{Z,p}_{n|n}) < \delta)$ and the resulting privacy-utility tradeoff; larger values of beta are expected to result in higher utility losses with lower probabilities of privacy outage while smaller values of beta are expected to result in lower utility losses with higher probabilities of privacy outage. The privacy-utility tradeoff region corresponding to different values of $\beta$ will be determined experimentally. 

We now use the Bellman equation to formulate the finite horizon privacy-utility trade-off optimization problem. Let $V_k$ denote the state-value function at timestep $k$ and $V_k^\ast(s_k)$ denote the optimal state-value function given the state $s_k = \{z_k, \widehat{x}^z_{k-1|k-1}, \widehat{x}^{\tilde{z}}_{k-1|k-1}, P^z_{k-1|k-1}, P^{\tilde{z}}_{k-1|k-1}\}$. Then, using the Bellman equation of optimality, the optimization problem can be formulated as: 

\begin{align*}
    V_k^\ast(s_k) 
    = \max_{C_k} \int_{\mathscr{S}}
        \prob(s_{k+1}|s_k, C_k)\cdot \\ 
        \Big( R_{k}(s_k, C_k, s_{k+1}) + V_{k+1}^\ast(s_{k+1}) \Big) \mathrm{d}s_{k+1}
    \\
    = \max_{C_k}
        \int_{\mathscr{Z}} \prob(z_{k+1}|s_k, C_k) \int_{\Lambda} \prob(\widehat{x}^z_{k|k}|s_k, C_k) \cdot \\
        \int_{\Delta} \prob(\widehat{x}^{\tilde{z}}_{k|k}|s_k, C_k) \int_{\Phi} \prob(P^z_{k|k}|s_k, C_k) \cdot \\
        \int_{\Omega} 
                \prob(P^{\tilde{z}}_{k|k}|s_k, C_k) \Big( R_{k}(s_k, C_k, s_{k+1}) + V_{k+1}^\ast(s_{k+1}) \Big) \cdot\\
                \, \mathrm{d}P^{\tilde{z}}_{k|k}  \, \mathrm{d}P^z_{k|k} \mathrm{d}\widehat{x}^{\tilde{z}}_{k|k} \, \mathrm{d}\widehat{x}^z_{k|k} \, \mathrm{d}z_{k+1},
\end{align*}
where $\mathscr{Z}, \Lambda, \Delta, \Phi$ and $\Omega$ are the feasible spaces of $z_{k+1}, \widehat{x}^z_{k|k}, \widehat{x}^{\tilde{z}}_{k|k}, P^z_{k|k}$ and $P^{\tilde{z}}_{k|k}$, respectively.

Given $s_k$ and $C_k$, the state variables, $\widehat{x}^z_{k|k}$, $\widehat{x}^{\tilde{z}}_{k|k}$, $P^z_{k|k}$ and $P^{\tilde{z}}_{k|k}$, are all deterministic (which directly follows from the application of the Kalman filter equations). Therefore,

\begin{align}
    V_k^\ast(s_k) 
    &= \max_{C_k} \int_{\mathscr{Z}}
    \!\begin{aligned}[t] 
        &\prob(z_{k+1}|s_k, C_k) \cdot \\
        &\Big( R_{k}(s_k, C_k, s_{k+1}) + V_{k+1}^\ast(s_{k+1}) \Big) \\ &\mathrm{d}z_{k+1}
    \end{aligned}
    \\
    &= \max_{C_k} \int_{\mathscr{Z}}
    \!\begin{aligned}[t] 
        &\prob(z_{k+1}|z_k) \cdot \\
        &\Big( R_{k}(s_k, C_k, s_{k+1}) + V_{k+1}^\ast(s_{k+1}) \Big) \\ &\mathrm{d}z_{k+1}. \label{eqn:bellman}
    \end{aligned}
\end{align}



\vspace{2mm}
If $X_k$ is a Gaussian process and $V_k$ is a Gaussian white noise process, then, $Z_{k+1}|Z_k \sim \mathrm{N}(\Bar{\mu}, \Bar{\Sigma})$ where
\begin{align*}
    \Bar{\mu} &= \EXPECT[Z_{k+1}|Z_k] \\
        &= \EXPECT[HX_{k+1}+V_{k+1} | Z_k] \\
        &= \EXPECT[HX_{k+1} | Z_k] + \EXPECT[V_{k+1} | Z_k] \\
        &= H \EXPECT[X_{k+1} | Z_k] \\
        &= H \widehat{x}^z_{k+1|k} \\
        &= HF \widehat{x}^z_{k|k}
\end{align*}
and
\begin{align*}
    \Bar{\Sigma} &= \COV(Z_{k+1}|Z_k) \\
        &= \COV(HX_{k+1}+V_{k+1} | Z_k) \\
        &= \COV[HX_{k+1} | Z_k] + \COV[V_{k+1} | Z_k] \\
        &= H \COV(X_{k+1}|Z_k) H^T + R \\
        &= H P^z_{k+1|k} H^T + R \\
        &= H (FP^z_{k|k}F^T +Q) H^T + R.
\end{align*}

The optimization problem in (\ref{eqn:bellman}) reflects the user's objective of maximizing the expected sum of the current and future rewards starting at a particular state $s_k$. The optimizing argument $C_k^\ast(s_k) = \argmax_{C_k} V_k^\ast(s_k)$ constitutes the best action taken toward the goal of maximizing the expected sum of rewards.


\section{Sub-optimal Algorithms} \label{section:sub_optimal}
The optimization problem formulated in (\ref{eqn:bellman}) suffers from the curse of dimensionality as both the state space and the action space are continuous. Analytical methods to solve the problem are infeasible for practical problems as they do not yield closed-form solutions for higher dimensional problems. Numerical algorithms, such as the \textit{value iteration} algorithm and the \textit{policy iteration} algorithm which advance by sweeping through all possible states at each time step, also fail as there are infinitely many states to sweep through. The problem is therefore not easily tractable without further assumptions about the state space or the action space, or both.

A popular approach to solving similar optimization problems involves discretizing the state space (see, for instance, \cite{uther1998tree, rust1997comparison, koenig1998xavier, iversen2014optimal, feng2012dynamic}). This approach is typically suboptimal, however, it can still yield promising solutions. In what follows, we highlight different algorithms that are based on the discretization of the state space to solve the optimization problem formulated in (\ref{eqn:bellman}).

\subsection{Value Iteration with Discretization}
The value iteration approach to solving a finite-horizon discrete state space MDP problem involves solving the Bellman equation to find the optimal values for every possible state at every time step, starting at the end of the finite time horizon and working backwards. At time step $n$, the optimal value of each state is computed using the terminal reward given by $R_n(S_n, C_n) = \beta d^p(n,n) - d^u(n,n)$. The algorithm then iteratively calculates the optimal values at previous time steps as given in Algorithm \ref{algorithm:value_iteration}.



\begin{algorithm}
\caption{Value Iteration Algorithm with Discretization}
\label{algorithm:value_iteration}
\begin{algorithmic}[1]
\State Define the feasible state space, $\mathscr{S}$.
\State Select a discretization rule, $\mathscr{D}$, and discretize $\mathscr{S}$ according to $\mathscr{D}$.  
\Procedure{Value Iteration}{}
\State Initialize $V_n^\ast(s_n)$ for all $s_n \in \mathscr{S}$ with terminal rewards.
\For{$k = n-1$ to $1$}
\For{each $s_k \in \mathscr{S}$}
\State $V_k^\ast(s_k)
    \!\begin{aligned}[t] 
        &= \max_{C_k} \sum_{z_{k+1}} \prob(z_{k+1}|z_k) \cdot \\
        &\Big( R_{k}(s_k, C_k, s_{k+1}) + V_{k+1}^\ast(s_{k+1}) \Big)
    \end{aligned}
    $
\State $C_k^\ast(s_k) = \argmax_{C_k} V_k^\ast(s_k)$
\EndFor
\EndFor
\EndProcedure
\end{algorithmic}
\end{algorithm}

The major characteristic of the value iteration algorithm is that it calculates optimal values and the optimal actions associated with the states at all time steps before the actual observations are available. The calculated values are optimal for the discrete MDP, however, due to the additional discretization step (which is not intrinsic to the value iteration algorithm itself), they are typically sub-optimal for the original MDP. 

\subsection{Pessimistic algorithm}
The pessimistic algorithm is a customized algorithm to solve the finite-horizon discrete state space MDP problem. The pessimistic algorithm captures an agent who always expects to transition to the worst possible state at every time step. The pessimistic algorithm seeks to optimize the value and the action associated with a state while assuming that the next transition leads to the state with the least value. The advantage of using the algorithm is that it is computationally less intensive as the state transition in the underlying model is assumed to be deterministic. The pessimistic algorithm is highlighted in Algorithm \ref{algorithm:pessimistic}.

\begin{algorithm}
\caption{Pessimistic Algorithm}
\label{algorithm:pessimistic}
\begin{algorithmic}[1]
\State Define the feasible state space, $\mathscr{S}$.
\State Select a discretization rule, $\mathscr{D}$, and discretize $\mathscr{S}$ according to $\mathscr{D}$.  
\State Initialize $V_n^\ast(s_n)$ for all $s_n \in \mathscr{S}$ with terminal rewards.
\State $v_n^{\#} = \min \{ V_n^\ast(s_n) \,:\, s_n \in \mathscr{S} \}$
\State $s_n^{\#} = \argmin_{s_n } \{ V_n^\ast(s_n) \,:\, s_n \in \mathscr{S} \}$
\For{$k = n-1$ to $1$}
\For{each $s_k \in \mathscr{S}$}
\State $V_k^\ast(s_k) = 
    \!\begin{aligned}[t] 
        &\max_{C_k} \Big( R_{k}(s_k, C_k, s^{\#}_{k+1}) + v_{k+1}^{\#} \Big)
    \end{aligned}
    $
\State $C_k^\ast(s_k) = \argmax_{C_k} V_k^\ast(s_k)$
\EndFor
\State $v_k^{\#} = \min \{ V_k^\ast(s_k) \,:\, s_k \in \mathscr{S} \}$
\State $s_k^{\#} = \argmin_{s_k} \{ V_k^\ast(s_k) \,:\, s_k \in \mathscr{S} \}$
\EndFor
\end{algorithmic}
\end{algorithm}

A quick remark on the notations: $\min \{ . \}$ represents the minimum of the set whereas $\argmin_Y \{ . \}$ represents the parameter $Y$ that corresponds to the minimum value of the set.

\subsection{Optimistic algorithm}
In contrast to the pessimistic algorithm, the optimistic algorithm captures an agent who always expects to transition to the best possible state at every time step. The optimistic algorithm seeks to optimize the value and the action associated with a state while assuming that the next transition leads to the state with the highest value. The optimistic algorithm is highlighted in Algorithm \ref{algorithm:optimistic}.

\begin{algorithm}
\caption{Optimistic Algorithm}
\label{algorithm:optimistic}
\begin{algorithmic}[1]
\State Define the feasible state space, $\mathscr{S}$.
\State Select a discretization rule, $\mathscr{D}$, and discretize $\mathscr{S}$ according to $\mathscr{D}$. 
\State Initialize $V_n^\ast(s_n)$ for all $s_n \in \mathscr{S}$ with terminal rewards.
\State $v_n^{\ast} = \max \{ V_n^\ast(s_n) \,:\, s_n \in \mathscr{S} \}$
\State $s_n^{\ast} = \argmax_{s_n} \{ V_n^\ast(s_n) \,:\, s_n \in \mathscr{S} \}$
\For{$k = n-1$ to $1$}
\For{each $s_k \in \mathscr{S}$}
\State $V_k^\ast(s_k) = 
    \!\begin{aligned}[t] 
        &\max_{C_k} \Big( R_{k}(s_k, C_k, s^{\ast}_{k+1}) + v_{k+1}^{\ast} \Big)
    \end{aligned}
    $
\State $C_k^\ast(s_k) = \argmax_{C_k} V_k^\ast(s_k)$
\EndFor
\State $v_k^{\ast} = \max \{ V_k^\ast(s_k) \,:\, s_k \in \mathscr{S} \}$
\State $s_k^{\ast} = \argmax_{s_k} \{ V_k^\ast(s_k) \,:\, s_k \in \mathscr{S} \}$
\EndFor
\end{algorithmic}
\end{algorithm}

\section{Privacy-Utility Tradeoff Under Estimated Privacy Leakage} \label{section:estimated_leakage}
The finite-horizon privacy-utility tradeoff problem can also be formulated with the constraint on estimated privacy leakage instead of the actual privacy leakage at the end of the finite time horizon. The resulting optimization problem is much simpler to solve, nevertheless, the dynamic privacy requirements are still captured into the problem formulation. To this end, we consider a user who seeks to maximize her instantaneous utility while limiting the estimated future leakage about her sensitive information. The resulting optimization problem is:

\begin{align}
    \min_{C_k}\ d(\widehat{X}^{Z,u}_{k|k}, \widehat{X}^{\tilde{Z},u}_{k|k}) \\
    \mbox{subject to:} \nonumber \\
    d(\widehat{X}^{Z,p}_{n|k}, \widehat{X}^{\tilde{Z},p}_{n|k}) \geq \delta \label{eqn:max_delta_constraint}
\end{align}

Intuitively, the user seeks to disclose as much as possible (allowed by the privacy constraint) at the current time step so as to maximize her current utility with a complete disregard for her future utilities. This strategy is, therefore, referred to as a \textit{maximum disclosure strategy}. In contrast, the optimization problem formulated in (\ref{eqn:bellman}) captures a user who seeks to cautiously disclose her personal information \textit{piecewise}. 

In a dynamic setting, a user following the maximum disclosure strategy needs to solve the optimization problem at every time step as new observations are made. As the user approaches the end of the finite time horizon, the privacy constraint is more restrictive due to the accumulated leakage resulting from the past disclosures. In some cases, the actual leakage may already exceed the estimated leakage and therefore, no choice of $C_k$ may satisfy the constraint, especially closer to the end of the finite time horizon. Therefore, it is more appropriate to formulate an unconstrained optimization problem that captures the semantics of the constrained problem. This leads us to the following optimization problem:

\begin{align}
    &\ \ \ \ \min_{C_k}\ d(\widehat{X}^{Z,u}_{k|k}, \widehat{X}^{\tilde{Z},u}_{k|k}) - \beta \left(
    d(\widehat{X}^{Z,p}_{n|k}, \widehat{X}^{\tilde{Z},p}_{n|k}) \right)
    \nonumber \\
    &= \max_{C_k} \beta \left(
    d(\widehat{X}^{Z,p}_{n|k}, \widehat{X}^{\tilde{Z},p}_{n|k}) \right) - d(\widehat{X}^{Z,u}_{k|k}, \widehat{X}^{\tilde{Z},u}_{k|k}) \label{eqn:max_unconstrained}
\end{align}

The parameter, $\beta$, in (\ref{eqn:max_unconstrained}) directly relates to the constraint in (\ref{eqn:max_delta_constraint}) and influences the probability of privacy outage at the end of the finite time horizon, $\prob(d(\widehat{X}^{\tilde{Z},p}_{n|n}, \widehat{X}^{Z,p}_{n|n}) < \delta)$. 

Although the optimization problem formulated in (\ref{eqn:max_unconstrained}) can easily be solved without further transformation, it is nevertheless possible to transform the optimization problem into an equivalent MDP formulation. First, define
\begin{align*}
   &d^u(k,k) \triangleq d(\widehat{X}^{Z,u}_{k|k}, \widehat{X}^{\tilde{Z},u}_{k|k}),
   \\
   &d^p(n,k) \triangleq d(\widehat{X}^{Z,p}_{n|k}, \widehat{X}^{\tilde{Z},p}_{n|k}),
   \\
   &S_k \triangleq \{Z_k\},
\end{align*}
where $S_k$ represents the state of the MDP. We now define the reward function, $R_k$, as
\begin{align*}
   R_k(S_k, C_k) = \beta d^p(n,k) - d^u(k,k)
\end{align*}

Notice that the reward function is independent of the future observations and therefore, deterministic. At any given time $k$, the user's goal is to chose an action $C_k = f(S_k)$ that allows her to maximize the instantaneous reward, $R_k(S_k, C_k)$. Since the user is oblivious to future rewards, we set $\gamma = 0$. The MDP equivalent of the optimization problem in (\ref{eqn:max_unconstrained}) can then be expressed as:

\begin{align}
    V_k^\ast(s_k) = \max_{C_k} R_k(s_k,C_k)   \label{eqn:max_disclosure_mdp}
\end{align}

The optimization problem in (\ref{eqn:max_disclosure_mdp}) reflects the user's objective of maximizing her instantaneous reward at a particular state $s_k$. The argument of the optimization $C_k^\ast(s_k) = \argmax_{C_k} V_k^\ast(s_k)$ constitutes the best action taken toward the goal of maximizing the instantaneous reward.

The main advantage of the optimization problem formulated in (\ref{eqn:max_disclosure_mdp}) (and equivalently, in \ref{eqn:max_unconstrained}) is that it does not require sweeping through all possible states at each time step (which would otherwise be required if the current reward depended on future states) and therefore, computationally much less intensive to solve. Further, the optimization problem is solved forwards as new observations become available. Algorithm \ref{algorithm:max_disclosure} highlights the steps involved in solving the finite-horizon privacy-utility tradeoff optimization problem under estimated privacy leakage using the maximum disclosure strategy.

\begin{algorithm}
\caption{Maximum Disclosure Algorithm}
\label{algorithm:max_disclosure}
\begin{algorithmic}[1]
\State At each time step $k$, \textbf{do}
\State \quad $V_k^\ast(s_k) = \max_{C_k} R_{k}(s_k, C_k)$
\State \quad $C_k^\ast(s_k) = \argmax_{C_k} V_k^\ast(s_k)$
\State \textbf{end}
\end{algorithmic}
\end{algorithm}

\section{Simulations} \label{section:simulations}

\begin{figure*}
 \centering
    \begin{subfigure}{0.48\textwidth}
    \centering
	\includegraphics[width=1.0\columnwidth]{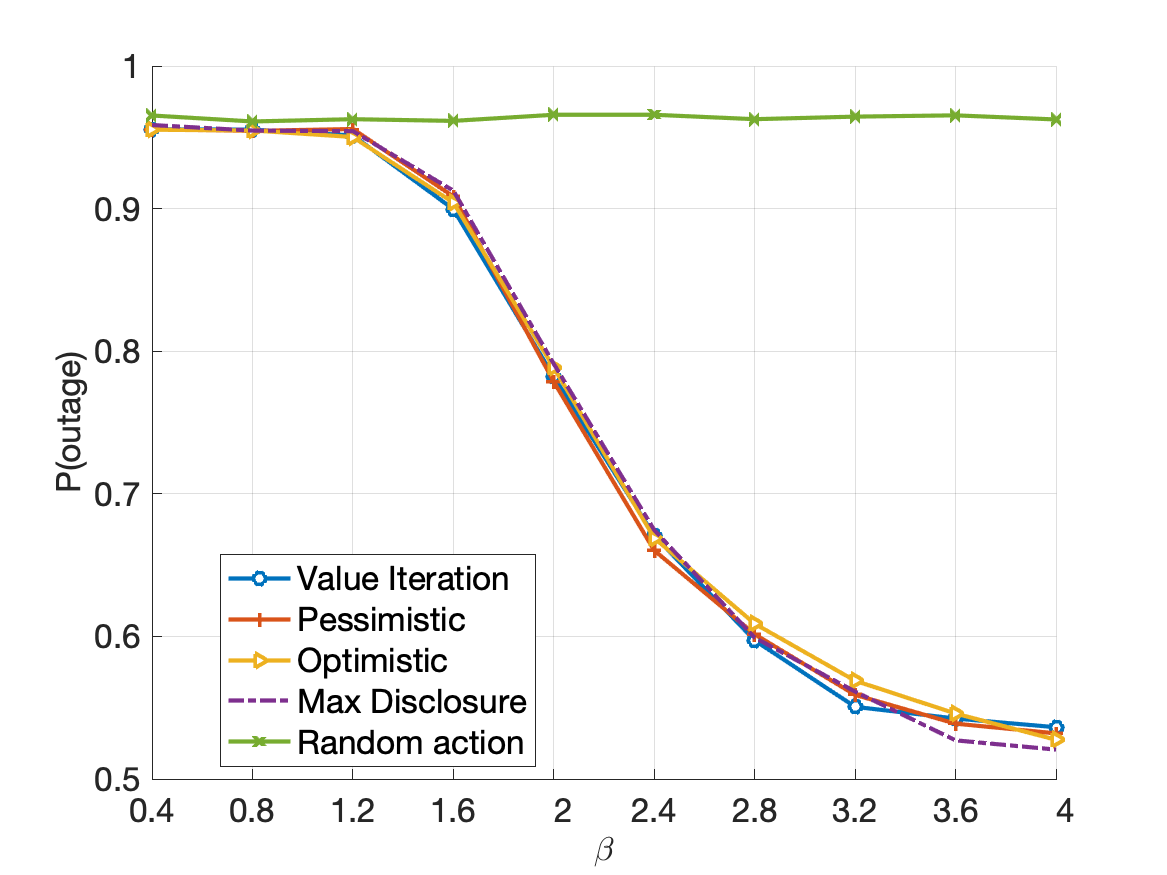}
	\caption{Probability of privacy outage for different values of $\beta$}
	\label{fig:poutage_n_eq_5}
    \end{subfigure}
    \hfill
    \begin{subfigure}{0.48\textwidth}
    \centering
	\includegraphics[width=1.0\columnwidth]{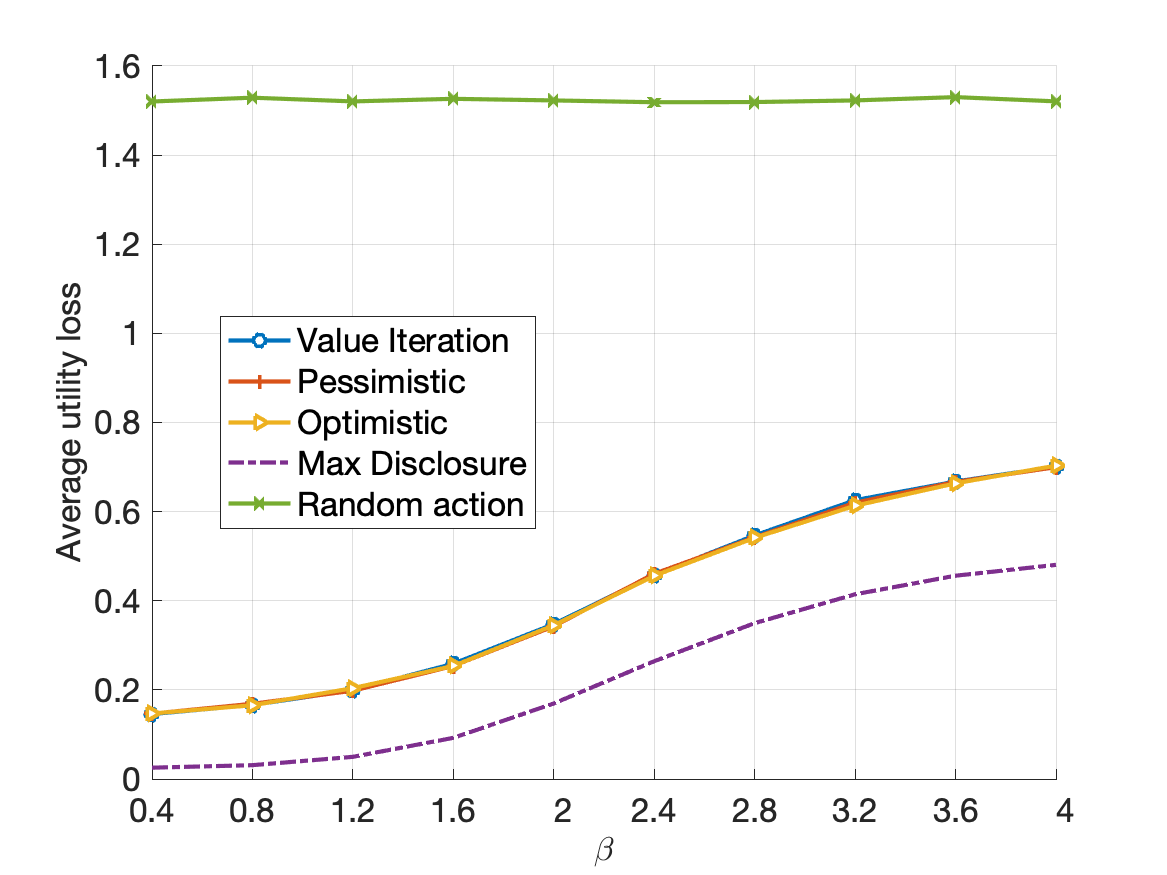}
	\caption{Average utility loss for different values of $\beta$}
	\label{fig:avguloss_n_eq_5}
    \end{subfigure}
    \caption{Comparison of the performances of different algorithms in terms of the probability of outage and the average utility loss for system model 1 ($\delta = 0.3$, $n = 5$ and $M = 1$).}
    \label{fig:performance_n_eq_5}
    \vspace{0mm}
\end{figure*}

\begin{figure}
  \centering
    \begin{minipage}{0.48\textwidth}
	\includegraphics[width=1.0\columnwidth]{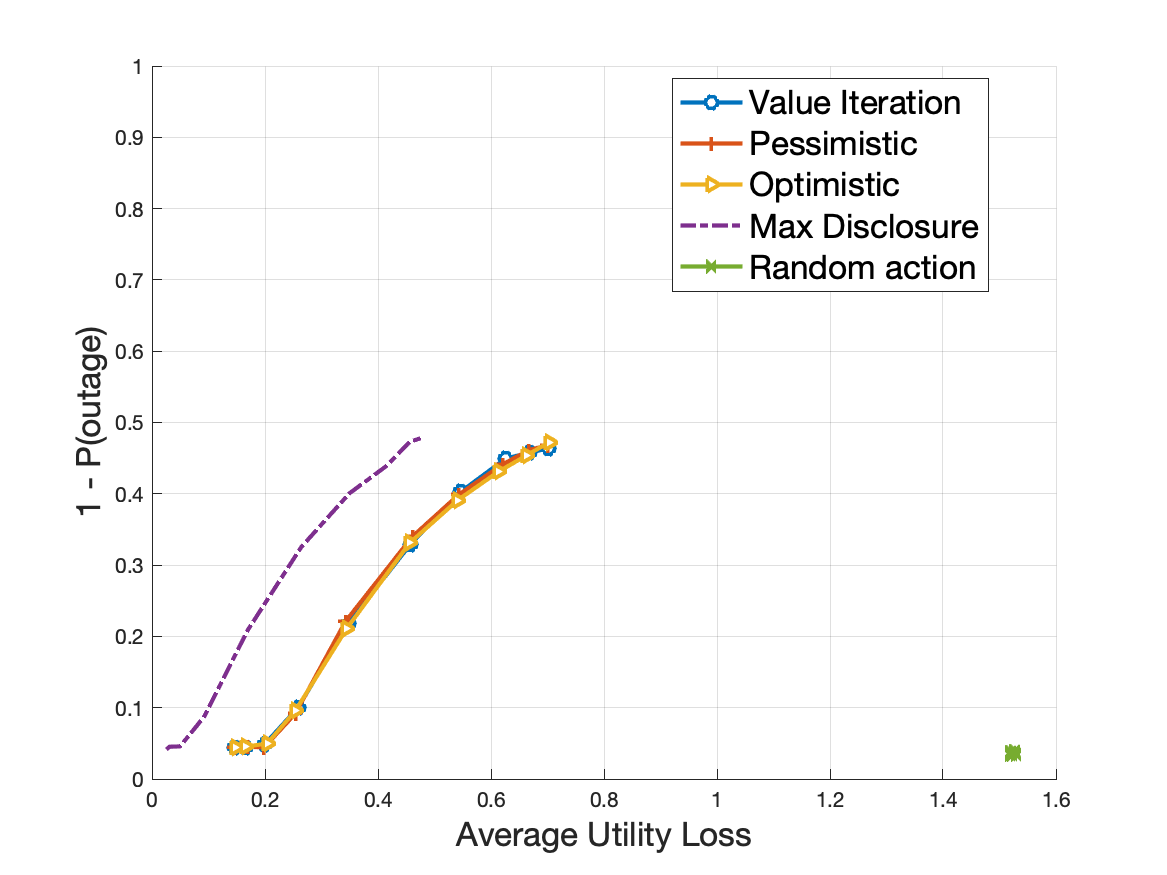}
    \caption{Privacy-utility tradeoff achieved by different strategies for system model 1 ($\delta = 0.3$, $n = 5$ and $M = 1$).}
    \label{fig:tradeoff_1}
    \end{minipage}
    \begin{minipage}{0.48\textwidth}
	\includegraphics[width=1.0\columnwidth]{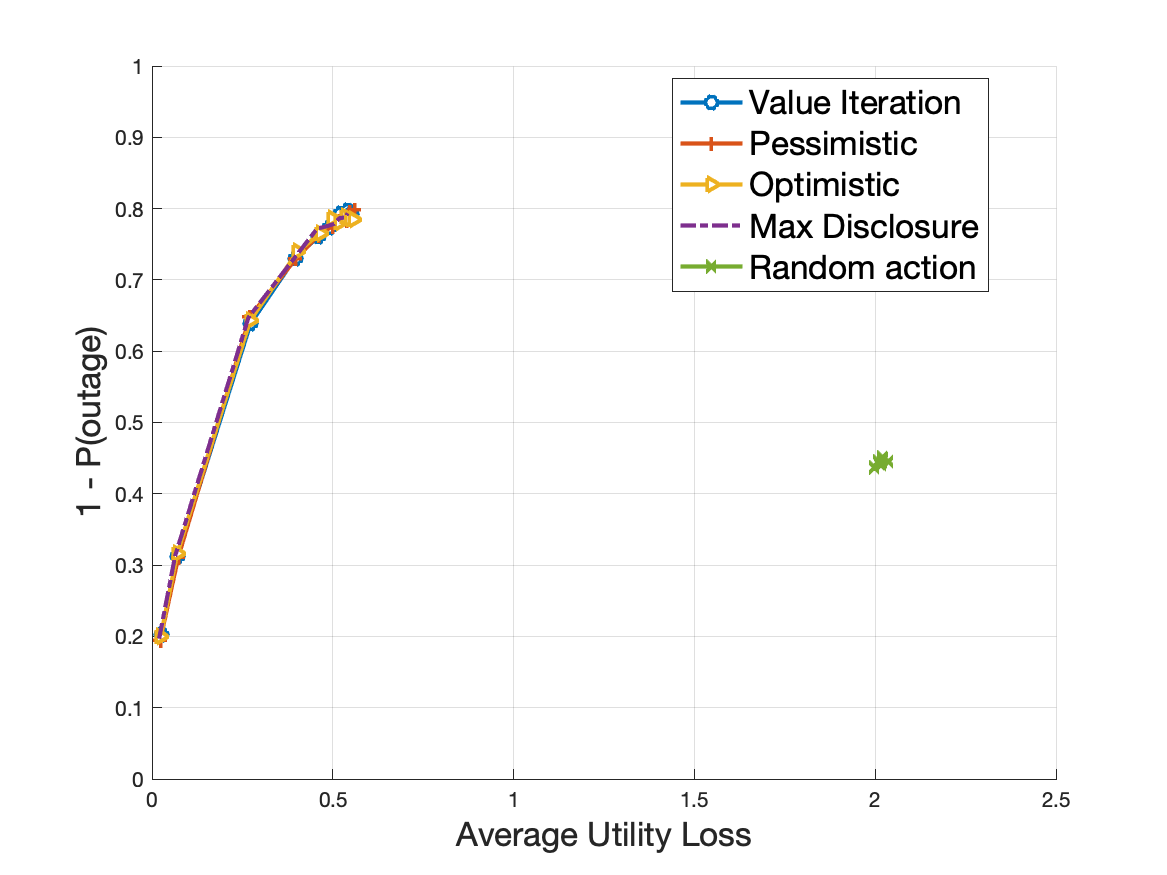}
    \caption{Privacy-utility tradeoff achieved by different strategies for system model 2 ($\delta = 0.3$, $n = 5$ and $M = 1$).}
    \label{fig:tradeoff_2}
    \end{minipage}
    \begin{minipage}{0.48\textwidth}
	\includegraphics[width=1.0\columnwidth]{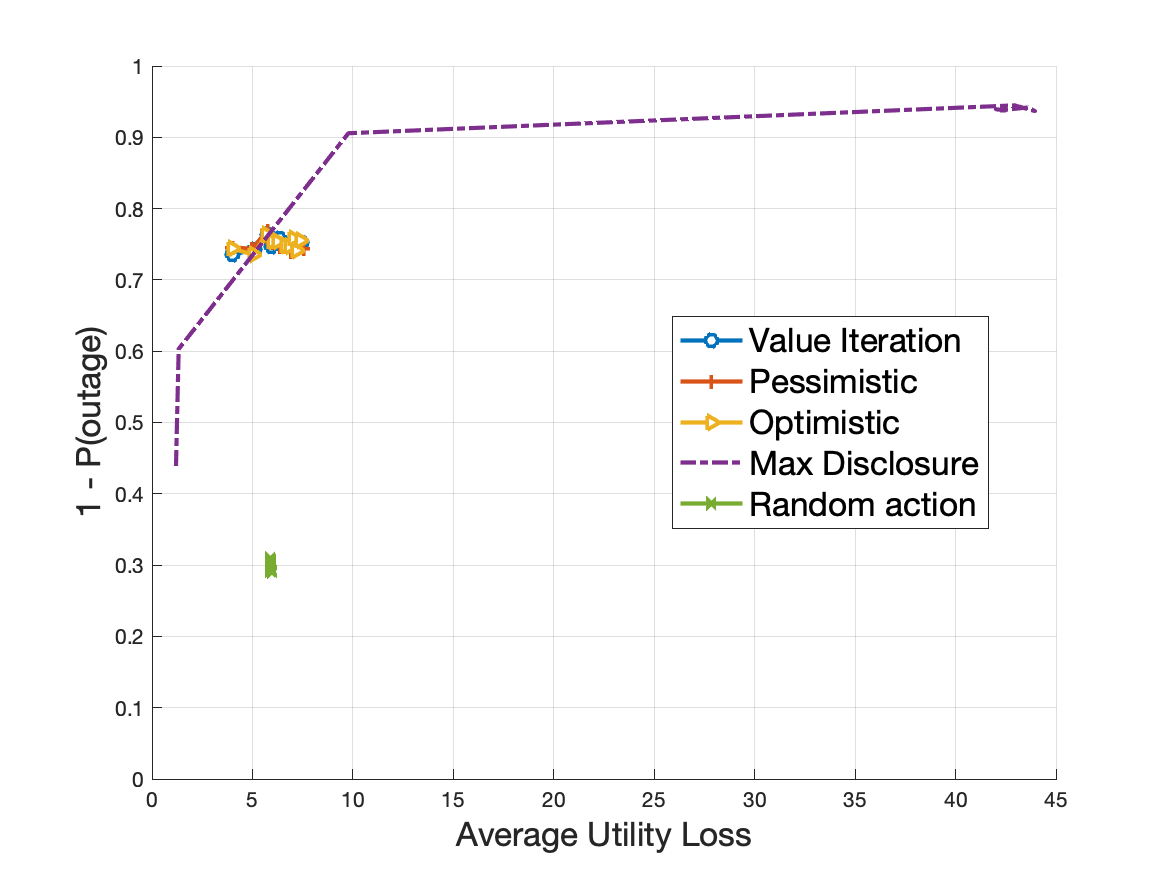}
    \caption{Privacy-utility tradeoff achieved by different strategies for system model 3 ($\delta = 0.3$, $n = 5$ and $M = 1$).}
    \label{fig:tradeoff_3}
    \end{minipage}
\end{figure}

In this section, we evaluate the performance of the value iteration algorithm, the pessimistic algorithm, the optimistic algorithm and the maximum disclosure algorithm via synthetic simulations. For our simulations, we consider an LDS with $N_p = 1$ and $N_u = 2$. We assume that $F_k$ and $H_k$ are time invariant such that $F_1 = F_2 = \cdots = F_n = F$ and $H_1 = H_2 = \cdots = H_n = H$ . Further, we assume that $X_k$ is a zero mean Gaussian process and $W_k$ and $V_k$ are independent and identically distributed standard Gaussian random vectors.

The elements of $F$ and $H$ are sampled independently from a uniform distribution in the unit interval. $F$ is further normalized such that its eigenvalues lie within a unit circle which ensures that the LDS is stable. As $P^Z_{k-1|k-1}$ is not a function of the observations or the actions, it is computed offline. Similarly, $P^{\tilde{Z}}_{k-1|k-1}$ is estimated with $P^{\tilde{Z}}_{k-1|0}$. For the value iteration, pessimistic and optimistic algorithms, a discretization function, $\mathscr{D}$, is used to approximate the components of  $Z_k$, $\widehat{X}^Z_{k-1|k-1}$ and $\widehat{X}^{\tilde{Z}}_{k-1|k-1}$ as binary variables. As a simplest discretization strategy, we chose the function $\mathscr{D}$ such that:
\begin{align}
    &\mathscr{D}(y) =
         &\begin{cases}    
            \EXPECT[y] - 0.1 & \text{when $y < \EXPECT[y]$,} \\
            \EXPECT[y] + 0.1 & \text{when $y \geq \EXPECT[y]$.} \label{eqn:discretization}
        \end{cases}
\end{align}
The choice of $0.1$ as the distance to the quantization points from the mean is arbitrary.

The performances of the four algorithms are evaluated in terms of the probability of privacy outage and the average utility loss. First, each algorithm, with the exception of the maximum disclosure algorithm, is run in turn to determine the optimal actions associated with every discretized state of the Markov Decision Process. Next, $10,000$ Monte-Carlo simulations of the LDS are carried out. In each simulation of the LDS, a sequence of observations, $z_1, z_2, \cdots, z_n$  are generated. When an observation $z_k$ is generated, the Kalman filter equations are used to compute the actual state, $s_k$. For the value iteration, the pessimistic and the optimistic algorithms, the Bellman equation (\ref{eqn:bellman}) is solved to determine the optimal action, $C_k^{\ast}$, associated with the state. For the maximum disclosure algorithm, the non-recursive equation (\ref{eqn:max_disclosure_mdp}) is solved to determine the optimal action, $C_k^{\ast}$, associated with the state. This process is repeated until the end of the finite time horizon. At the end of the finite time horizon, any violation of the privacy constraint: $d(\widehat{X}^{\tilde{Z},p}_{n|n}, \widehat{X}^{Z,p}_{n|n}) < \delta$, is checked, which concludes one simulation. After all simulations have been completed, the probability of privacy outage and the average utility loss are calculated using
\begin{align*}
    &\prob(\text{outage}) = \frac{\text{number of constraint violations}}{\text{total number of simulations}} \\
    \text{and}\\
    &\text{Average utility loss} = \frac{\sum_{}^{} \big( \sum_{k=1}^{n} d(\widehat{X}^{Z,u}_{k|k}, \widehat{X}^{\tilde{Z},u}_{k|k}) \big)}{\text{total number of simulations}},
\end{align*}
respectively. The experiment is repeated multiple times for different randomly generated samples of $H$ and $F$.

From among multiple system models used in the simulations, three representative system models are selected that provide various insights on the performances of the four algorithms:

\begin{tcolorbox}
{\scriptsize
\noindent\textbf{Model 1:}
\begin{align*}
    F = 
    \begin{bmatrix}
    0.06218 & 0.08373 & 0.12324 \\
    0.07386 & 0.04809 & 0.11332 \\
    0.13481 & 0.09099 & 0.06936
    \end{bmatrix}, \\
    H = 
    \begin{bmatrix}
    0.30780 & 0.77969 & 0.29994 \\
    0.37514 & 0.67681 & 0.45616 \\
    0.98334 & 0.94292 & 0.45824
    \end{bmatrix}.
\end{align*}
}
\end{tcolorbox}

\begin{tcolorbox}
{\scriptsize
\noindent\textbf{Model 2:}
\begin{align*}
    F = 
    \begin{bmatrix}
    0.02712 & 0.01067 & 0.00073 \\
    0.00792 & 0.01444 & 0.01576 \\
    0.01029 & 0.00998 & 0.01596
    \end{bmatrix}, \\
    H = 
    \begin{bmatrix}
    0.02712 & 0.01067 & 0.00073 \\
    0.00792 & 0.01444 & 0.01576 \\
    0.01029 & 0.00998 & 0.01596
    \end{bmatrix}.
\end{align*}
}
\end{tcolorbox}

\begin{tcolorbox}
{\scriptsize
\noindent\textbf{Model 3:}
\begin{align*}
    F = 
    \begin{bmatrix}
    0.12246 & 0.51340 & 0.14024 \\
    0.45475 & 0.02484 & 0.53664 \\
    0.35442 & 0.70248 & 0.05728
    \end{bmatrix}, \\
    H = 
    \begin{bmatrix}
    0.75237 & 0.31551 & 0.85396 \\
    0.93524 & 0.03364 & 0.62274 \\
    0.01605 & 0.36138 & 0.05232
    \end{bmatrix}.
\end{align*}
}
\end{tcolorbox}

Figure \ref{fig:performance_n_eq_5} shows the performances of the four algorithms in terms of the probability of privacy outage and the average utility loss across different values of $\beta$ for system model 1. For reference, the performance of a na\"ive strategy in which the user randomly selects $C_k$ from a uniform distribution in the unit interval at each time step $k$ is also included. In Figure \ref{fig:poutage_n_eq_5} and Figure \ref{fig:avguloss_n_eq_5}, we observe that all four algorithms consistently outperform the \textit{random action} strategy across all values of $\beta$. Also, for all four algorithms, we observe a decrease in the probability of privacy outage, and an increase in the average utility loss, as $\beta$ increases. This observation is consistent with the intuition that larger values of $\beta$ put more weight on the privacy requirement than the utility requirement and therefore, result in a decrease in the probability of privacy outage and an increase in the utility loss. For the random action strategy, however, we observe that the probability of privacy outage (Figure \ref{fig:poutage_n_eq_5}) and the average utility loss (Figure \ref{fig:avguloss_n_eq_5}) are both virtually constant across all values of $\beta$. This is expected as the random action strategy does not account for $\beta$ in the selection of $C_k$.

In Figure \ref{fig:poutage_n_eq_5} and Figure \ref{fig:avguloss_n_eq_5}, we also observe that the performances of the value iteration, the pessimistic and the optimistic algorithms are similar across different values of $\beta$. We consistently observed similar performances of the three algorithms for different random samples of $H$ and $F$ and for different values of $n$. In light of this, we conclude that the average performances of the three algorithms in terms of the probability of outage and the average utility loss are similar. Consequently, it may be desirable to use the pessimistic or the optimistic algorithm over the value iteration algorithm for speed benefits. However, it should be noted that the algorithms may perform differently for a more robust discretization strategy.

Figure \ref{fig:tradeoff_1}, Figure \ref{fig:tradeoff_2} and Figure \ref{fig:tradeoff_3} highlight the privacy-utility tradeoff achieved by the maximum disclosure strategy/algorithm against the value iteration, pessimistic and optimistic algorithms for the three system models. For system model 1 (Figure \ref{fig:tradeoff_1}), we observe that the maximum disclosure strategy significantly outperforms the other algorithms and achieves much better privacy-utility tradeoff. For system model 2 (Figure \ref{fig:tradeoff_2}), we observe that the performance of the maximum disclosure strategy is similar to that of the value iteration, pessimistic and optimistic algorithms. Similarly, for system model 3 (Figure \ref{fig:tradeoff_3}), the dynamic range for the privacy-utility tradeoff achieved by the maximum disclosure strategy is significantly higher than the other strategies-- the higher dynamic range translates to more room for tuning the privacy-utility tradeoff.

The performance of the four algorithms, in general, depends on the system model. The relatively poor performances of the value iteration, the pessimistic and the optimistic algorithms against the maximum disclosure strategy across all representative system models can be attributed to the choice of the binary discretization strategy. For a more robust discretization strategy, we expect the three algorithms to perform better than the maximum disclosure strategy. However, increasing the quantization points in pursuit of a better discretization strategy significantly increases the computational requirements and may not be feasible for all systems.  For high dimensional practical problems, maximum disclosure strategy is therefore the only computationally feasible option to solve the dynamic privacy problem.

\section{Conclusion} \label{section:conclusion}
The increasing privacy concerns associated with disclosing personal data have guided the state-of-the-art privacy mechanisms that afford theoretical privacy guarantees. However, existing works mostly consider the problem in a static setting which either assumes that a user discloses her information only once, or treats each disclosure of the user's information independently to the previous disclosures. In this paper, we considered a dynamic model of privacy in which a privacy-aware user cautiously discloses her personal information to a service provider over a finite time horizon. We investigated different strategies that allow a user to maximize their utility over time while limiting the privacy leakage at the end of the finite horizon. Using experimental evaluations on synthetic datasets, we showed that these strategies, although sub-optimal, can yield promising tradeoff region for the finite-horizon privacy-utility tradeoff problem. Finally, we demonstrated that there exists a simpler strategy corresponding to a simplified version of the problem, that has significant computational benefits with encouraging performance.




%

\printbibliography

@article{sankar2013utility,
  title={Utilitear-privacy tradeoffs in databases: An information-theoretic approach},
  author={Sankar, Lalitha and Rajagopalan, S Raj and Poor, H Vincent},
  journal={IEEE Transactions on Information Forensics and Security},
  volume={8},
  number={6},
  pages={838--852},
  year={2013},
  publisher={IEEE}
}

@article{zhu2014correlated,
  title={Correlated differential privacy: Hiding information in non-IID data set},
  author={Zhu, Tianqing and Xiong, Ping and Li, Gang and Zhou, Wanlei},
  journal={IEEE Transactions on Information Forensics and Security},
  volume={10},
  number={2},
  pages={229--242},
  year={2014},
  publisher={IEEE}
}

@article{farokhi2017fisher,
  title={Fisher information as a measure of privacy: Preserving privacy of households with smart meters using batteries},
  author={Farokhi, Farhad and Sandberg, Henrik},
  journal={IEEE Transactions on Smart Grid},
  volume={9},
  number={5},
  pages={4726--4734},
  year={2017},
  publisher={IEEE}
}

@article{wu2017game,
  title={Game theory based correlated privacy preserving analysis in big data},
  author={Wu, Xiaotong and Wu, Taotao and Khan, Maqbool and Ni, Qiang and Dou, Wanchun},
  journal={IEEE Transactions on Big Data},
  year={2017},
  publisher={IEEE}
}

@inproceedings{makhdoumi2013privacy,
  title={Privacy-utility tradeoff under statistical uncertainty},
  author={Makhdoumi, Ali and Fawaz, Nadia},
  booktitle={2013 51st Annual Allerton Conference on Communication, Control, and Computing (Allerton)},
  pages={1627--1634},
  year={2013},
  organization={IEEE}
}

@inproceedings{rajagopalan2011smart,
  title={Smart meter privacy: A utility-privacy framework},
  author={Rajagopalan, S Raj and Sankar, Lalitha and Mohajer, Soheil and Poor, H Vincent},
  booktitle={2011 IEEE international conference on smart grid communications (SmartGridComm)},
  pages={190--195},
  year={2011},
  organization={IEEE}
}

@inproceedings{alvim2011differential,
  title={Differential privacy: on the trade-off between utility and information leakage},
  author={Alvim, M{\'a}rio S and Andr{\'e}s, Miguel E and Chatzikokolakis, Konstantinos and Degano, Pierpaolo and Palamidessi, Catuscia},
  booktitle={International Workshop on Formal Aspects in Security and Trust},
  pages={39--54},
  year={2011},
  organization={Springer}
}

@article{braun2009quantitative,
  title={Quantitative notions of leakage for one-try attacks},
  author={Braun, Christelle and Chatzikokolakis, Konstantinos and Palamidessi, Catuscia},
  journal={Electronic Notes in Theoretical Computer Science},
  volume={249},
  pages={75--91},
  year={2009},
  publisher={Elsevier}
}

@inproceedings{smith2009foundations,
  title={On the foundations of quantitative information flow},
  author={Smith, Geoffrey},
  booktitle={International Conference on Foundations of Software Science and Computational Structures},
  pages={288--302},
  year={2009},
  organization={Springer}
}

@inproceedings{li2009tradeoff,
  title={On the tradeoff between privacy and utility in data publishing},
  author={Li, Tiancheng and Li, Ninghui},
  booktitle={Proceedings of the 15th ACM SIGKDD international conference on Knowledge discovery and data mining},
  pages={517--526},
  year={2009},
  organization={ACM}
}

@inproceedings{issa2016operational,
  title={An operational measure of information leakage},
  author={Issa, Ibrahim and Kamath, Sudeep and Wagner, Aaron B},
  booktitle={2016 Annual Conference on Information Science and Systems (CISS)},
  pages={234--239},
  year={2016},
  organization={IEEE}
}

@article{li2018maximal,
  title={Maximal correlation secrecy},
  author={Li, Cheuk Ting and El Gamal, Abbas},
  journal={IEEE Transactions on Information Theory},
  volume={64},
  number={5},
  pages={3916--3926},
  year={2018},
  publisher={IEEE}
}

@inproceedings{asoodeh2015maximal,
  title={On maximal correlation, mutual information and data privacy},
  author={Asoodeh, Shahab and Alajaji, Fady and Linder, Tam{\'a}s},
  booktitle={2015 IEEE 14th Canadian Workshop on Information Theory (CWIT)},
  pages={27--31},
  year={2015},
  organization={IEEE}
}

@inproceedings{wang2017estimation,
  title={An estimation-theoretic view of privacy},
  author={Wang, Hao and Calmon, Flavio P},
  booktitle={2017 55th Annual Allerton Conference on Communication, Control, and Computing (Allerton)},
  pages={886--893},
  year={2017},
  organization={IEEE}
}

@inproceedings{erdogdu2015privacy,
  title={Privacy-utility trade-off under continual observation},
  author={Erdogdu, Murat A and Fawaz, Nadia},
  booktitle={2015 IEEE International Symposium on Information Theory (ISIT)},
  pages={1801--1805},
  year={2015},
  organization={IEEE}
}

@inproceedings{kalantari2017information,
  title={On information-theoretic privacy with general distortion cost functions},
  author={Kalantari, Kousha and Sankar, Lalitha and Kosut, Oliver},
  booktitle={2017 IEEE International Symposium on Information Theory (ISIT)},
  pages={2865--2869},
  year={2017},
  organization={IEEE}
}

@article{yin2017location,
  title={Location privacy protection based on differential privacy strategy for big data in industrial internet of things},
  author={Yin, Chunyong and Xi, Jinwen and Sun, Ruxia and Wang, Jin},
  journal={IEEE Transactions on Industrial Informatics},
  volume={14},
  number={8},
  pages={3628--3636},
  year={2017},
  publisher={IEEE}
}

@inproceedings{yang2012differential,
  title={Differential privacy in data publication and analysis},
  author={Yang, Yin and Zhang, Zhenjie and Miklau, Gerome and Winslett, Marianne and Xiao, Xiaokui},
  booktitle={Proceedings of the 2012 ACM SIGMOD International Conference on Management of Data},
  pages={601--606},
  year={2012},
  organization={ACM}
}

@inproceedings{farokhi2017optimal,
  title={Optimal privacy-preserving policy using constrained additive noise to minimize the fisher information},
  author={Farokhi, Farhad and Sandberg, Henrik},
  booktitle={2017 IEEE 56th Annual Conference on Decision and Control (CDC)},
  pages={2692--2697},
  year={2017},
  organization={IEEE}
}

@inproceedings{wang2018preserving,
  title={PRESERVING PARAMETER PRIVACY IN SENSOR NETWORKS},
  author={Wang, Chong Xiao and Song, Yang and Tay, Wee Peng},
  booktitle={2018 IEEE Global Conference on Signal and Information Processing (GlobalSIP)},
  pages={1316--1320},
  year={2018},
  organization={IEEE}
}

@inproceedings{liao2017hypothesis,
  title={Hypothesis testing under maximal leakage privacy constraints},
  author={Liao, Jiachun and Sankar, Lalitha and Calmon, Flavio P and Tan, Vincent YF},
  booktitle={2017 IEEE International Symposium on Information Theory (ISIT)},
  pages={779--783},
  year={2017},
  organization={IEEE}
}

@inproceedings{basciftci2016privacy,
  title={On privacy-utility tradeoffs for constrained data release mechanisms},
  author={Basciftci, Yuksel Ozan and Wang, Ye and Ishwar, Prakash},
  booktitle={2016 Information Theory and Applications Workshop (ITA)},
  pages={1--6},
  year={2016},
  organization={IEEE}
}

@inproceedings{he2014blowfish,
  title={Blowfish privacy: Tuning privacy-utility trade-offs using policies},
  author={He, Xi and Machanavajjhala, Ashwin and Ding, Bolin},
  booktitle={Proceedings of the 2014 ACM SIGMOD international conference on Management of data},
  pages={1447--1458},
  year={2014}
}

@inproceedings{sharma2021practical,
  title={A Practical Approach to Navigating the Tradeoff Between Privacy and Precise Utility},
  author={Sharma, Chandra and Mandal, Bishwas and Amariucai, George},
  booktitle={ICC 2021-IEEE International Conference on Communications},
  pages={1--6},
  year={2021},
  organization={IEEE}
}

@INPROCEEDINGS{8462600,  
author={Song, Yang and Wang, Chong Xiao and Tay, Wee Peng},  
booktitle={2018 IEEE International Conference on Acoustics, Speech and Signal Processing (ICASSP)},
title={Privacy-Aware Kalman Filtering},   
year={2018},  
volume={},  
number={}, 
pages={4434-4438},  
doi={10.1109/ICASSP.2018.8462600}}

@ARTICLE{8768395,  author={Song, Yang and Wang, Chong Xiao and Tay, Wee Peng},  journal={IEEE Transactions on Information Forensics and Security},   title={Compressive Privacy for a Linear Dynamical System},   year={2020},  volume={15},  number={},  pages={895-910},  doi={10.1109/TIFS.2019.2930366}}

@article{welch2020kalman,
  title={Kalman filter},
  author={Welch, Gregory F},
  journal={Computer Vision: A Reference Guide},
  pages={1--3},
  year={2020},
  publisher={Springer}
}

@inproceedings{zhou2009continuous,
  title={Continuous privacy preserving publishing of data streams},
  author={Zhou, Bin and Han, Yi and Pei, Jian and Jiang, Bin and Tao, Yufei and Jia, Yan},
  booktitle={Proceedings of the 12th International Conference on Extending Database Technology: Advances in Database Technology},
  pages={648--659},
  year={2009}
}

@inproceedings{dwork2010differential,
  title={Differential privacy under continual observation},
  author={Dwork, Cynthia and Naor, Moni and Pitassi, Toniann and Rothblum, Guy N},
  booktitle={Proceedings of the forty-second ACM symposium on Theory of computing},
  pages={715--724},
  year={2010}
}

@article{kung2017compressive,
  title={Compressive privacy: From information$\backslash$/estimation theory to machine learning [lecture notes]},
  author={Kung, Sun-Yuan},
  journal={IEEE Signal Processing Magazine},
  volume={34},
  number={1},
  pages={94--112},
  year={2017},
  publisher={IEEE}
}

@inproceedings{diamantaras2016data,
  title={Data privacy protection by kernel subspace projection and generalized eigenvalue decomposition},
  author={Diamantaras, Konstantinos and Kung, Sun-Yuan},
  booktitle={2016 IEEE 26th International Workshop on Machine Learning for Signal Processing (MLSP)},
  pages={1--6},
  year={2016},
  organization={IEEE}
}

@article{kung2018compressive,
  title={A compressive privacy approach to generalized information bottleneck and privacy funnel problems},
  author={Kung, Sun-Yuan},
  journal={Journal of the Franklin Institute},
  volume={355},
  number={4},
  pages={1846--1872},
  year={2018},
  publisher={Elsevier}
}

@article{kung2017collaborative,
  title={Collaborative PCA/DCA learning methods for compressive privacy},
  author={Kung, Sun-Yuan and Chanyaswad, Thee and Chang, J Morris and Wu, Peiyuan},
  journal={ACM Transactions on Embedded Computing Systems (TECS)},
  volume={16},
  number={3},
  pages={1--18},
  year={2017},
  publisher={ACM New York, NY, USA}
}

@article{he2013distortion,
  title={A distortion-based approach to privacy-preserving metering in smart grids},
  author={He, Xingze and Zhang, Xinwen and Kuo, C-C Jay},
  journal={IEEE Access},
  volume={1},
  pages={67--78},
  year={2013},
  publisher={IEEE}
}

@inproceedings{liao2018privacy,
  title={Privacy under hard distortion constraints},
  author={Liao, Jiachun and Kosut, Oliver and Sankar, Lalitha and Calmon, Flavio P},
  booktitle={2018 IEEE Information Theory Workshop (ITW)},
  pages={1--5},
  year={2018},
  organization={IEEE}
}

@article{kalantari2018robust,
  title={Robust privacy-utility tradeoffs under differential privacy and hamming distortion},
  author={Kalantari, Kousha and Sankar, Lalitha and Sarwate, Anand D},
  journal={IEEE Transactions on Information Forensics and Security},
  volume={13},
  number={11},
  pages={2816--2830},
  year={2018},
  publisher={IEEE}
}

@article{wang2016relation,
  title={On the relation between identifiability, differential privacy, and mutual-information privacy},
  author={Wang, Weina and Ying, Lei and Zhang, Junshan},
  journal={IEEE Transactions on Information Theory},
  volume={62},
  number={9},
  pages={5018--5029},
  year={2016},
  publisher={IEEE}
}

@inproceedings{dwork2006our,
  title={Our data, ourselves: Privacy via distributed noise generation},
  author={Dwork, Cynthia and Kenthapadi, Krishnaram and McSherry, Frank and Mironov, Ilya and Naor, Moni},
  booktitle={Annual International Conference on the Theory and Applications of Cryptographic Techniques},
  pages={486--503},
  year={2006},
  organization={Springer}
}

@article{geng2015optimal,
  title={The optimal noise-adding mechanism in differential privacy},
  author={Geng, Quan and Viswanath, Pramod},
  journal={IEEE Transactions on Information Theory},
  volume={62},
  number={2},
  pages={925--951},
  year={2015},
  publisher={IEEE}
}

@inproceedings{koufogiannis2017differential,
  title={Differential privacy for dynamical sensitive data},
  author={Koufogiannis, Fragkiskos and Pappas, George J},
  booktitle={2017 IEEE 56th Annual Conference on Decision and Control (CDC)},
  pages={1118--1125},
  year={2017},
  organization={IEEE}
}

@article{sugiura2021bayesian,
  title={Bayesian Differential Privacy for Linear Dynamical Systems},
  author={Sugiura, Genki and Ito, Kaito and Kashima, Kenji},
  journal={IEEE Control Systems Letters},
  year={2021},
  publisher={IEEE}
}

@article{han2018privacy,
  title={Privacy in control and dynamical systems},
  author={Han, Shuo and Pappas, George J},
  journal={Annual Review of Control, Robotics, and Autonomous Systems},
  volume={1},
  pages={309--332},
  year={2018},
  publisher={Annual Reviews}
}

@article{uther1998tree,
  title={Tree based discretization for continuous state space reinforcement learning},
  author={Uther, William TB and Veloso, Manuela M},
  journal={Aaai/iaai},
  volume={98},
  pages={769--774},
  year={1998}
}

@article{rust1997comparison,
  title={A comparison of policy iteration methods for solving continuous-state, infinite-horizon Markovian decision problems using random, quasi-random, and deterministic discretizations},
  author={Rust, John P},
  journal={Infinite-Horizon Markovian Decision Problems Using Random, Quasi-random, and Deterministic Discretizations (April 1997)},
  year={1997}
}

@article{koenig1998xavier,
  title={Xavier: A robot navigation architecture based on partially observable markov decision process models},
  author={Koenig, Sven and Simmons, Reid and others},
  journal={Artificial Intelligence Based Mobile Robotics: Case Studies of Successful Robot Systems},
  number={partially},
  pages={91--122},
  year={1998},
  publisher={Citeseer}
}

@article{iversen2014optimal,
  title={Optimal charging of an electric vehicle using a Markov decision process},
  author={Iversen, Emil B and Morales, Juan M and Madsen, Henrik},
  journal={Applied Energy},
  volume={123},
  pages={1--12},
  year={2014},
  publisher={Elsevier}
}

@article{feng2012dynamic,
  title={Dynamic programming for structured continuous Markov decision problems},
  author={Feng, Zhengzhu and Dearden, Richard and Meuleau, Nicolas and Washington, Richard},
  journal={arXiv preprint arXiv:1207.4115},
  year={2012}
}

\end{document}